\documentclass[10pt,conference]{IEEEtran}

\usepackage{xspace}
\usepackage{multirow}
\usepackage{booktabs}
\usepackage{xcolor}
\usepackage{graphicx}
\usepackage[caption=false,font=footnotesize]{subfig}
\usepackage{tabularx}
\usepackage{makecell}
\usepackage{mdframed}
\usepackage{float}
\usepackage{comment}
\usepackage{url}
\usepackage[hidelinks]{hyperref}

\definecolor{lightgreenbg}{HTML}{E8F5E9}
\definecolor{darkgreenline}{HTML}{2E7D32}

\newcounter{finding}
\newcommand{\finding}[1]{%
    \refstepcounter{finding}%
    \vspace{1mm}%
    \begin{mdframed}[
        linecolor=darkgreenline,
        roundcorner=0pt,
        backgroundcolor=lightgreenbg,
        linewidth=3pt,
        innerleftmargin=5pt,
        leftmargin=0cm,
        rightmargin=0cm,
        topline=false,
        bottomline=false,
        rightline=false
    ]
        \textbf{Finding \arabic{finding}:} #1
    \end{mdframed}
    \vspace{1mm}%
}

\AtBeginDocument{%
  }

\newcommand{\ours}{DAIRA\xspace}
\newcommand{\ds}{DeepSeek V3.2\xspace}
\newcommand{\mini}{Mini-SWE-agent\xspace}
\newcommand{\live}{Live-SWE-agent\xspace}
\newcommand{\geminiflash}{Gemini 3 Flash Preview\xspace}
\newcommand{\qwen}{Qwen-3-Coder Flash\xspace}
\newcommand{\openhand}{OpenHands\xspace}
\newcommand{\sweagent}{SWE-agent\xspace}
\newcommand{\tracetool}{Dynamic Analysis Tool\xspace}
\newcommand{\myworkflow}{Test-Tracing Driven Workflow\xspace}
\newcommand{\tracesummary}{Execution Trace Semantic Analysis\xspace}
\newcommand{\swebenchv}{SWE-bench Verified\xspace}
\newcommand{\swebenchm}{SWE-bench Multilingual\xspace}

\newcommand{\up}[1]{\rlap{\hspace{0.2em}\scriptsize\textcolor{red}{#1}}}
\newcommand{\down}[1]{\rlap{\hspace{0.2em}\scriptsize\textcolor{teal!60!black}{#1}}}

\title{DAIRA: Dynamic Analysis-enhanced Issue Resolution Agent}

\author{
\begin{tabular}{ccc}
\begin{tabular}{c}
Mingwei Liu\\
Sun Yat-sen University\\
Zhuhai, China\\
liumw26@mail.sysu.edu.cn
\end{tabular}
&
\begin{tabular}{c}
Zihao Wang\\
Sun Yat-sen University\\
Zhuhai, China\\
wangzh778@mail2.sysu.edu.cn
\end{tabular}
&
\begin{tabular}{c}
Zhenxi Chen\\
Sun Yat-sen University\\
Zhuhai, China\\
chenzhx236@mail2.sysu.edu.cn
\end{tabular}
\\[1.2em]
\begin{tabular}{c}
Zheng Pei\\
Sun Yat-sen University\\
Zhuhai, China\\
peizh3@mail2.sysu.edu.cn
\end{tabular}
&
\begin{tabular}{c}
Yanlin Wang\\
Sun Yat-sen University\\
Zhuhai, China\\
yanlin-wang@outlook.com
\end{tabular}
&
\end{tabular}
}

\begin{document}

\maketitle

\begin{abstract}
Large language models (LLMs) have recently enabled software agents to tackle repository-level issue resolution from natural language problem reports. Despite this progress, current agents still observe programs mostly through static source code. They therefore lack direct evidence about runtime behavior, which makes complicated failures hard to localize and can push agents toward wide-ranging, trial-and-error exploration with limited repair effectiveness.

This paper introduces \textbf{\ours}, a \emph{Dynamic Analysis-enhanced Issue Resolution Agent} designed to make execution behavior observable to LLM-based repair agents. \ours does not require agents to operate low-level debuggers or inspect noisy raw traces. Instead, it uses a lightweight tracing mechanism to capture execution information and convert it into structured execution trace reports. These reports summarize runtime evidence, including call paths, component roles, and state transitions, in a form that supports agent reasoning. As a result, agents can relate dynamic behavior to static code structure and make fault localization and patch generation more evidence-driven.

We evaluate \ours on \swebenchv for Python and on multilingual issue-resolution tasks covering Python, C/C++, Ruby, and Java. On \swebenchv, \ours reaches a 79.4\% resolution rate with \geminiflash, surpassing five agent-based frameworks that use different LLM backbones. Across all evaluated languages, \ours delivers a 13.32\% overall relative improvement over the SWE-agent baseline. The results show that structured execution-level observability can consistently improve repository-level issue resolution across programming languages.

\end{abstract}

\begin{IEEEkeywords}
Issue Resolution, Dynamic Analysis, LLM
\end{IEEEkeywords}

\section{Introduction}
\label{sec:introduction}
Issue resolution is a central yet challenging task in real-world software engineering~\cite{Tassey2002TheEI}. Modern software systems evolve through continuous user--developer interaction, where bugs are reported via natural language issue descriptions (e.g., GitHub Issues)~\cite{10.1145/1453101.1453146}. Resolving such issues requires developers to interpret informal descriptions, navigate large codebases, identify fault locations, and construct correct patches under complex execution semantics~\cite{jimenez2024swebenchlanguagemodelsresolve}. Consequently, automating issue resolution has become a long-standing goal in automated software engineering~\cite{NEURIPS2024_5a7c9475}.

Benchmarks such as SWE-bench have accelerated the development of LLM-based repair agents~\cite{wang2025openhandsopenplatformai,mu2025experepairdualmemoryenhancedllmbased}. Representative systems such as \sweagent~\cite{NEURIPS2024_5a7c9475} and its extensions rely on agent--environment interaction with static code context, often enhanced by search or experience-driven strategies~\cite{antoniades2025swesearchenhancingsoftwareagents,li-etal-2025-codetree,xia2025livesweagentsoftwareengineeringagents}. 
However, these methods primarily rely on static code context and lack \textbf{execution-level observability}, i.e., visibility into fine-grained program execution behavior such as call sequences and runtime state changes, which is critical for diagnosing complex defects. 

Existing agents construct context through static analysis and code retrieval, selecting relevant snippets based on lexical or semantic similarity. Without guidance from real execution evidence, this process can become speculative, often introducing irrelevant context while failing to expose the actual fault behavior. More critically, static code representations inherently fail to capture execution-specific behaviors. Real-world defects often arise from dynamic control flows, polymorphism, and implicit state mutations. Recent studies show that LLMs significantly degrade when reasoning over such scenarios~\cite{Liu2025EvaluatingCR}, indicating that static reasoning alone is insufficient for reliably capturing execution states required for debugging and repair.

In contrast, human developers rely on dynamic analysis, such as tracing and runtime inspection, to understand program behavior~\cite{2ed676bb99b54ce4afeac9ed977e89d6}. By exposing execution paths and intermediate states, dynamic analysis provides execution evidence that complements static reasoning. However, execution-level observability remains largely absent from current LLM-based repair agents~\cite{zhong2024debuglikehumanlarge,10.1145/3180155.3182533}.

Motivated by this gap, we argue that effective issue resolution requires execution-level observability to transform speculative reasoning into evidence-guided inference. To this end, we propose \textbf{\ours}, a \emph{\textbf{D}ynamic \textbf{A}nalysis--enhanced \textbf{I}ssue \textbf{R}esolution \textbf{A}gent} for repository-level repair. \ours adopts a trace-driven execution analysis paradigm that converts program executions into structured execution trace reports.

Specifically, \ours introduces a dynamic analysis module into the agent workflow. A lightweight tracer collects execution traces with fine-grained runtime events such as function calls and state transitions. These traces are then transformed into structured execution trace reports, which capture execution paths, component-level functional roles, and state changes. The agent performs reasoning over these reports to localize faults and generate patches.

We evaluate \ours on \swebenchv~\cite{jimenez2024swebenchlanguagemodelsresolve} and \swebenchm~\cite{yang2025swesmith}, together with trajectory- and cost-oriented analyses that study how dynamic execution information affects agent behavior. On \swebenchv, \ours achieves a \textbf{$79.4\%$} resolution rate with \geminiflash, outperforming prior agent-based methods. It also reaches \textbf{44.4\%} on more challenging issues. Across Python, C/C++, Ruby, and Java evaluations, \ours consistently improves over \sweagent, achieving an overall relative improvement of \textbf{13.32\%}. Further analyses show that dynamic execution information reduces unnecessary code exploration and improves patch quality across tasks.

Our main contributions are as follows:

\begin{itemize}
    \item \textbf{Language-agnostic execution tracing for LLM agents:} We design a lightweight and unified tracing interface that captures execution-level information in a model-friendly format, enabling consistent instrumentation across multiple programming languages.

    \item \textbf{Dynamic analysis-enhanced issue resolution agent:} We integrate dynamic execution signals into an LLM-based repair agent, enabling reasoning over runtime behaviors and improving fault localization and patch generation.

    \item \textbf{Empirical study on repository-level repair:} We evaluate \ours on \swebenchv and \swebenchm across multiple programming languages, showing consistent improvements and analyzing the impact of execution-level information on agent behavior.
\end{itemize}

\section{Motivating Examples}

\begin{figure*}[t] % 濡傛灉鏄弻鏍忚鏂囧苟甯屾湜鍗犳弧閫氭爮锛岃浣跨敤 figure*
    \centering

    \includegraphics[width=\linewidth]{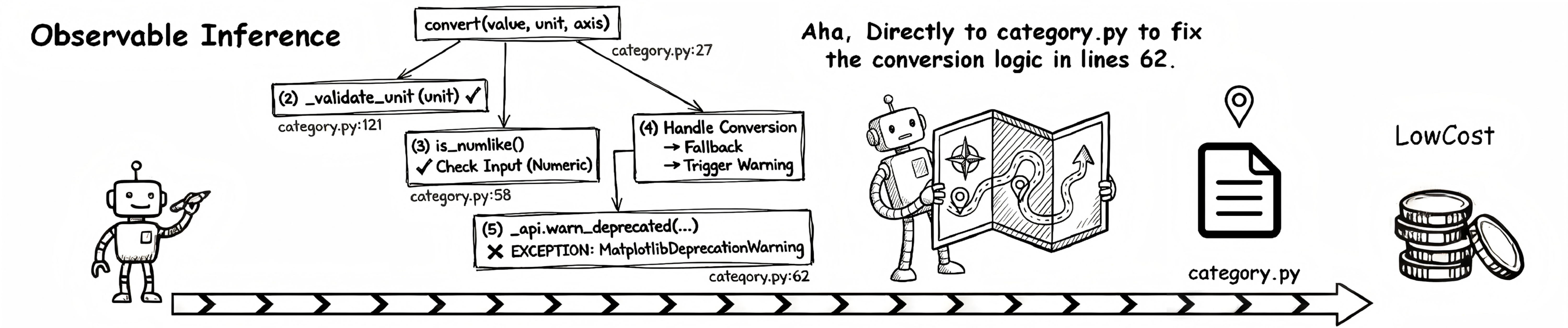}    
    \caption{\textbf{Impact of Dynamic Execution Traces on Debugging Trajectories.}} 
    \label{fig:motivation_main}
\end{figure*}

To demonstrate the effectiveness of dynamic execution information in automated issue resolution, we present two representative issues from Matplotlib~\cite{Hunter:2007} and SymPy~\cite{10.7717/peerj-cs.103}.

\begin{figure*}[t]
    \centering
    \includegraphics[width=0.96\textwidth,trim=0 18 0 18,clip]{images_example2.pdf}
    \vspace{-0.5em}
    \caption{\textbf{Comparison of Opaque and Transparent Runtime States.}}
    \label{fig:example2}
\end{figure*}

\textbf{Scenario 1: Trace-guided Context Retrieval for Precise Localization.}
Matplotlib-22719 describes a misleading DeprecationWarning triggered when passing empty data (\texttt{[]}) to axes with categorical units. The root cause lies in the \texttt{is\_numlike()} check within the \texttt{convert} function, which incorrectly classifies an "empty list" as numeric, thereby erroneously triggering an unrelated deprecation path during the conversion process. The main challenge of resolving this bug is the deceptive disconnect between the empty input and the numerical warning, causing agents to get lost in irrelevant call stacks.

Figure~\ref{fig:motivation_main} shows a comparison of the solution trajectories for the Matplotlib-22719 issue under different conditions. 
Due to the lack of visibility into the intermediate execution process, the agent falls into \textbf{Speculative Exploration}. {As shown in the upper part of Figure~\ref{fig:motivation_main}, the agent initiates a series of blind attempts.} 
Initially, it suspects an error in the upstream caller, questioning whether \texttt{axis.py} passed bad input to the \texttt{convert} function. Subsequently, it attempts to verify the internal state of \texttt{convert}, speculating whether the \texttt{\_validate\_unit} check passed or if the function itself is broken. To further interpret the conditional logic, it redundantly retrieves large unrelated files like \texttt{unit.py}. This speculative behavior compels the agent to load massive irrelevant code, leading to a ``High Context'' burden and significantly increasing repair costs.

Conversely, if the intermediate execution process is visible, the agent can directly pinpoint the root cause via \textbf{Observable Inference}. As depicted in the bottom path of Figure~\ref{fig:motivation_main}, the visualized execution process acts as a panoramic map. The agent directly observes that the error occurs because an empty list is incorrectly classified as numeric by \texttt{is\_numlike()}, triggering the erroneous deprecation warning. With this precise evidence, the agent bypasses redundant searches and navigates straight to \texttt{category.py} to repair the conversion.

\textbf{Scenario 2: Trace-Driven Accurate Detection for Complex Control Flows.}
SymPy-17630 describes a situation where, during block matrix multiplication and summation operations involving a zero matrix, the matrix object unexpectedly degenerates into scalar zero under \texttt{\_postprocessor} optimization paths, resulting in the loss of dimension properties and potentially causing system crashes. As Figure~\ref{fig:example2} shows, SymPy's complex, polymorphic call stack in block matrix computation makes isolating hidden type mismatches extremely challenging without precise localization.

Under the \textbf{opaque runtime state}, exploration is severely constrained, bounded entirely by the reported failure in \texttt{blockmatrix.py}. As shown on the left side of Figure~\ref{fig:example2}, matrix multiplication involves a highly complex computational process. Because unexpected type degenerations are hidden within this extensive codebase, the investigation scope drastically expands, leading to a severe \textbf{search space explosion}. Moreover, actual call paths are obscured by intricate program control flows and polymorphic operations (e.g., dynamic dispatch through \texttt{\_eval\_matrix\_mul} and \texttt{\_postprocessor}). With call targets dynamically determined at runtime, reconstructing exact execution traces through static inspection alone becomes unfeasible. In contrast, as illustrated in the right panel of Figure~\ref{fig:example2}, the \textbf{transparent runtime state} overcomes these debugging constraints via full call graph reconstruction. The availability of precise execution paths and intermediate visibility clearly delineates the entire execution flow. Such deep visibility makes it possible to immediately capture the critical anomaly and trace it directly to \texttt{matadd.py}, thereby seamlessly isolating the exact point of failure within the \texttt{rm\_id} rule.

\textbf{Design Inspiration: Integration of Dynamic Execution Information.}
These cases underscore a core limitation of static code analysis: defects rooted in 
dynamic control flow or implicit state changes often elude detection without runtime observation. Therefore, akin to human engineers using debuggers, equipping agents with intermediate state visualization is essential to strengthening their diagnostic capabilities.

\begin{comment}
    In summary, the limitations exposed in these cases reveal that relying solely on static code retrieval is insufficient for resolving defects buried within dynamic control flows or implicit state mutations. This observation motivates a fundamental shift in agent design: if we can equip the model with a dynamic analysis mechanism that perceives the code through actual execution rather than mere static reading, we have the potential to break through the "cognitive ceiling" of current agents. Such capability would enable agents to bridge the gap between static logic and runtime reality, effectively solving deep-seated logical defects that remain invisible to traditional text-based inference.
\end{comment}

\section{Approach}\label{sec:Approach}
Inspired by these examples, our goal is to equip agents with dynamic execution information. To this end, we introduce \ours, an automated issue resolution framework that 
encapsulates dynamic analysis as a callable tool, enabling the effective capture and analysis of intermediate execution states. Next, we detail the \tracetool and \tracesummary modules, along with our dynamic-analysis-tailored \myworkflow.
\subsection{\tracetool}
\begin{comment}
    {Recent efforts introduce dynamic analysis into LLM-based code tasks via block-based state tracking or breakpoint-driven debugging~\cite{}. However, these methods lack flexibility in complex scenarios. Specifically, block-based approaches exhibit poor generalizability, requiring tedious re-chunking for new codebases. Meanwhile, breakpoint interactions primarily target short-horizon tasks. In issue resolution demanding long-range reasoning, these granular actions trigger a surge in reasoning steps and a massive context burden. Moreover, the inherent complexity of these debugging tools hinders their deployment across the intricate environmental dependencies of real-world repositories.} 
\end{comment}
Integrating a dynamic analysis module into a long-horizon reasoning agent poses several inherent difficulties. In complex issue resolution, granular actions such as breakpoint interactions trigger a drastic proliferation of reasoning steps, which inevitably imposes an overwhelming context burden on the agent. Furthermore, achieving sufficient flexibility across diverse scenarios remains a significant hurdle; for instance, adapting block-based state tracking to new codebases necessitates a tedious re-chunking process due to its poor generalizability. Finally, the inherent intricacy of dynamic debugging tools substantially hinders deployment across the complex environmental dependencies typically found in real-world repositories. To address the above challenges, \ours adopts a lightweight ``trigger-and-collect'' tracing strategy. The \tracetool is integrated into the agent's action space via a standardized CLI and produces a unified structured trace format. In our Python backend, it is built on a customized version of the open-source Hunter engine~\cite{python_hunter}; for other language versions, we keep the same CLI and trace format while replacing the underlying backend with language-specific instrumentation, including compiler-based instrumentation for C/C++, JDWP/JDB or JFR for Java, and TracePoint for Ruby. As shown in Figure~\ref{fig:tracelog2report}, the tool captures the raw execution trace of the target script.
\begin{figure*}[htbp]
    \centering
    % 璇风‘淇濆浘鐗囨枃浠跺悕涓?trace_log.PDF锛屽苟涓?tex 鏂囦欢涓殑璺緞涓€鑷?    \includegraphics[width=\linewidth]{images_trace_log.pdf}
    \caption{Execution traces for reproducing Matplotlib issue \#22719.}
    \label{fig:tracelog2report}
\end{figure*}

\textbf{Native Execution via Automatic Hooks}. We describe this mechanism using the Python backend as an example. Forcing the agent to adhere to complex, fixed-format code injection imposes a high cognitive load, rendering the generation process highly error-prone and brittle. To mitigate this, we implement a Native In-Process Execution strategy. Upon invocation, it executes agent-synthesized standard Python reproduction scripts directly via the \texttt{runpy} interface. This format is consistent with running scripts, completely avoiding the need for agents to master specific debugging syntax or write boilerplate code. Guided by the specified parameters, the tool automatically injects low-level hooks leveraging \texttt{sys.settrace} into the runtime environment to capture process data. Although this dynamic instrumentation inherently introduces runtime overhead, such execution cost is largely marginal when compared to the latency of the agent's large language model (LLM) inference. Furthermore, by restricting the trace scope to task-relevant modules, we ensure that the execution remains efficient. Ultimately, this environment-agnostic design minimizes adaptation barriers, preserves the agent's workflow, and ensures high portability across legacy or complex Python environments.

\textbf{Spatiotemporal Trace Filtering}. To extract core execution insights while minimizing noise, we utilize two key parameters: trace scope and target function. Temporally, guided by the target function, the tracer adopts an "on-demand" approach---activating exclusively upon function invocation. Spatially, the trace scope acts as a whitelist, effectively filtering out extraneous libraries and dependencies. Regarding trace granularity, we selectively capture only function calls, returns, and exceptions. By restricting the trace to call-return pairs, this design achieves maximal trace compression while preserving the integrity of the entire execution trajectory.

\textbf{Adaptive Granularity Iteration}. Static configurations struggle to adapt to the dynamic complexity of defects. To prevent deep call stacks from triggering context explosion, we introduce a dynamic adjustment mechanism. When execution traces exceed the valid context window, an overflow signal triggers iterative correction, prompting the agent to reduce trace depth and rerun. By adjusting analysis granularity based on complexity, this mechanism completely avoids situations where excessive execution traces prevent dynamic analysis, enabling the module to dynamically adapt to different model context windows.

\subsection{\tracesummary}
\begin{figure}[htbp] % 鍘绘帀鏄熷彿锛屾敼涓哄崟鏍?    \centering
    % 灏嗗搴︽敼涓烘爮瀹斤紝寤鸿璁句负 0.9 鎴?1.0 鍊嶇殑 \columnwidth
    \includegraphics[width=0.95\columnwidth]{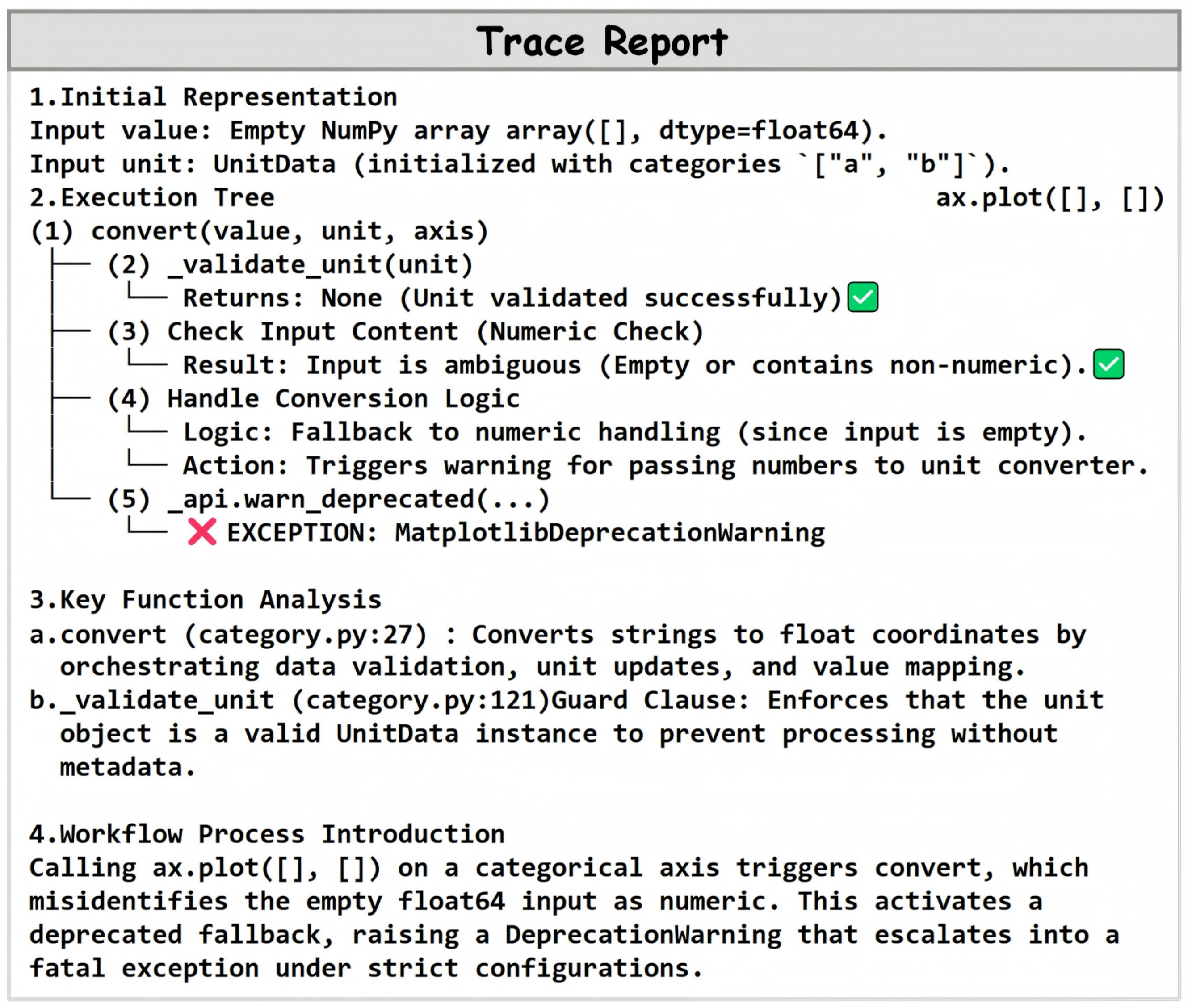}
    \caption{Execution trace report during the resolution process of Issue Matplotlib-22719.}
    \label{fig:tracereport1}
\end{figure}

To bridge the semantic gap between raw execution traces and high-level logic, \ours features a \tracesummary module. Following data collection by \tracetool, this module employs an LLM alongside the source code to analyze and reconstruct the raw execution traces into a transparent program execution path. The trace semantic analysis is performed by the same backbone LLM used by the repair agent, and its token usage and cost are included in the reported efficiency measurements. This synthesis yields a structured execution trace report (Figure~\ref{fig:tracereport1}) with three components.

\textbf{Hierarchical Logic Reconstruction:} Unlike a simple semantic restatement, this module organizes runtime behavior into a visual ASCII execution tree. As a standard software engineering format, ASCII execution trees use minimal tokens and visual indentation to explicitly map nested calls and data states. This efficiently depicts complex runtime control flows, fully unlocking LLMs' pre-trained topological reasoning capabilities. As shown in Figure~\ref{fig:tracereport1}, the execution tree leverages hierarchical indentation to structure complex runtime flows, thereby explicitly mapping conditional branches and pinpointing the exact exception triggers.

\textbf{Key Function Analysis:} By leveraging the call-return pairs explicitly recorded in execution traces, key function analysis elucidates the "workflow role" of each component to assist agents in understanding the codebase's function specifications. In Figure~\ref{fig:tracereport1}, the analysis clarifies that \texttt{convert} orchestrates the data validation, while \texttt{\_validate\_unit} acts as a guard clause to enforce valid \texttt{UnitData} instances. To prevent cascading anomalies during remediation, this module explicitly defines the functional boundaries of key components. This empowers agents to deduce component responsibilities by analyzing their role variations across different input scenarios.

\begin{figure*}[t]
    \centering
    \includegraphics[width=0.96\textwidth]{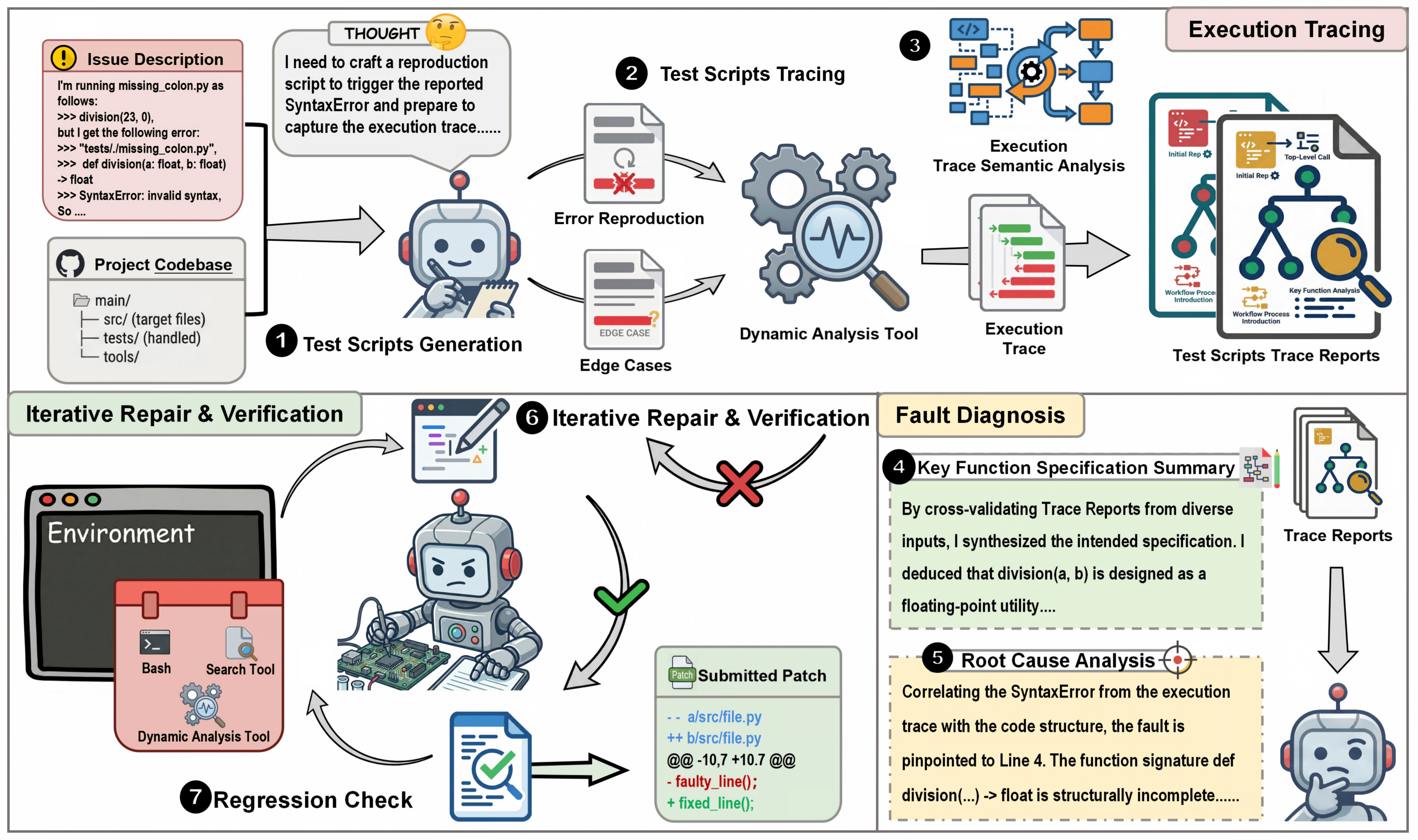}
    \caption{Overview of \ours{}'s \myworkflow.}
    \label{fig:7-steps-workflow}
\end{figure*}

\textbf{Workflow Process Introduction:} This section provides a natural language overview of the execution events. As the report shows, it summarizes low-level code behavior into high-level intent descriptions, thus supplementing the agent's understanding of the underlying trace data. Crucially, this external, objective description acts as a corrective mechanism against agent-authored test confirmation bias, ensuring feedback aligns with actual runtime behavior.
By integrating execution trees, key function analysis, and workflow descriptions, this module presents the agent with a ``panoramic view'' of execution. 

\subsection{\myworkflow}
Figure~\ref{fig:7-steps-workflow} shows an example of an ideal process. \ours adopts the \myworkflow not as a rigid procedural pipeline, but rather as a flexible behavioral strategy for autonomous agents. This strategy is specifically designed to accommodate the newly integrated dynamic analysis module, empowering agents to capture sufficient execution traces to guide the entire repair process. Our complete code and prompts are publicly available~\cite{daira_code_2025}.

\noindent\textbf{Phase 1: Execution Tracing Phase.} In this phase, the primary objective is to accumulate sufficient and diverse dynamic evidence at the beginning of the repair process. Compared with typical end-to-end repair agents, where test execution is primarily used as a post-patch validation signal, \ours shifts the agent's autonomous test-based validation to this initial stage. The agent generates reproduction and scenario scripts following the standard repair-agent setting, using the issue description, repository context, and available project tests. This validation step remains autonomous and iterative: when a script is invalid or insufficient, the agent revises or adds scripts based on runtime feedback. The validated scripts are then traced by the \textbf{\tracetool}, and the resulting raw execution traces are transformed by \textbf{\tracesummary} into structured execution trace reports.

\noindent\textbf{Phase 2: Fault Diagnosis Phase.} Next, the agent uses the gathered execution trace reports to map runtime anomalies back to the static code structure. It first advances through the \textbf{Key Function Specification Summary} stage to distill the overarching design intent and functional boundaries. Guided by this established intent, the workflow seamlessly transitions into \textbf{Root Cause Analysis} to pinpoint the exact components deviating from their expected logic. Ultimately, this systematic progression empowers the agent to move beyond superficial defensive fixes, enabling deep systemic corrections.

\noindent\textbf{Phase 3: Repair \& Verification Phase.} Finally, this phase ensures repair quality through a closed-loop verification process. Guided by the identified root cause, the agent enters the \textbf{Iterative Repair \& Verification} stage, modifying the source codebase and conducting dynamic validation within the environment. Empowered to autonomously re-invoke \tracetool, the agent assesses the applied modifications and clarifies any ambiguous logic. Should the validation fail, it iteratively refines the patch. Before finalizing the solution, a \textbf{Regression Check} is executed to ensure the repair resolves the targeted issue without introducing unintended side-effects, thereby guaranteeing the integrity of the codebase.

\section{Evaluation}
\label{sec:Evaluation}
To systematically evaluate the performance of \ours{} across multiple dimensions, our empirical study addresses the following research questions (RQs):

\noindent\textbf{RQ1: How does the performance of \ours compare to existing SOTA baselines in issue resolution tasks?}

\noindent\textbf{RQ2: Can the effectiveness gains of \ours extend beyond Python to other programming languages?}

\noindent\textbf{RQ3: What is the impact of different foundation models on the performance and efficiency of \ours?}

\noindent\textbf{RQ4: How does \ours perform across issues of varying difficulty levels?}

\noindent\textbf{RQ5: What is the impact of different components on the performance of \ours?}

\subsection{Experiment Setup}
\noindent\textbf{Benchmark.} We evaluate \ours on \swebenchv~\cite{jimenez2024swebenchlanguagemodelsresolve}, a widely regarded benchmark for automated issue resolution. It is a human-filtered subset of 500 instances from SWE-bench, created with OpenAI to ensure clear problem descriptions, correct test patches, and solvable tasks. We also use \swebenchm~\cite{yang2025swesmith}, which extends SWE-bench to 300 curated tasks across 9 programming languages, including C, C++, Go, and other languages.

\noindent\textbf{Baselines.} To evaluate the effectiveness of our proposed method, we select several SOTA methods as baselines, representing the most advanced and widely recognized approaches in the field of automated software engineering:

\textbf{\sweagent}~\cite{NEURIPS2024_5a7c9475}: A pioneering framework that introduced the Agent-Computer Interface (ACI) to optimize LLM interactions with software environments.

\textbf{\live}~\cite{xia2025livesweagentsoftwareengineeringagents}: A self-evolving extension of \mini that synthesizes custom tools at runtime, currently holding the SOTA position among open-source agents.

\textbf{\mini}~\cite{NEURIPS2024_5a7c9475}: A cost-effective, lightweight variant of \sweagent designed for streamlined reasoning.

\textbf{\openhand}~\cite{wang2025openhandsopenplatformai}: A flexible open-source platform supporting diverse agentic workflows through terminal commands.

\noindent\textbf{Implementation.} We implemented our approach based on \sweagent due to its stability and widespread adoption in the open-source community. Our method is framework-agnostic and can be adapted to agent frameworks with a configured runtime environment. 

\subsection{RQ1: Effectiveness}
\begin{table}[htbp]
  \centering
  \caption{Resolution performance of \ours and other advanced methods}
  \label{tab:performance}
  \renewcommand{\arraystretch}{1.08}
  \begin{tabular}{llc}
    \toprule
    \textbf{Method} & \textbf{Base Model} & \textbf{Resolved (\%)} \\
    \midrule

    \multirow{2}{*}{\live} & Claude Sonnet 4.5 & 75.40 \\
     & Gemini 3 Pro Preview & 77.40 \\
    \midrule
    \multirow{3}{*}{\openhand} & Claude Opus 4.5 & 77.60 \\
     & DeepSeek V3.2 & 71.60 \\
     & Gemini 3 Flash Preview & 74.60 \\
    \midrule
    \multirow{3}{*}{\mini} & Claude 4.5 Opus & 76.80 \\
     & \ds & 70.00\\
     & \geminiflash & 75.80\\
    \midrule
    \multirow{3}{*}{\sweagent} & Claude Sonnet 4.5 & 69.80 \\
     & \ds & 65.40 \\
     & \geminiflash & 73.60 \\
    \midrule
    \multirow{2}{*}{\textbf{\ours}} & \ds & 74.20 \\
     & \geminiflash & \textbf{79.40} \\
    \bottomrule
  \end{tabular}
\end{table}

\begin{figure}[htbp]
    \centering
    \includegraphics[width=0.62\columnwidth]{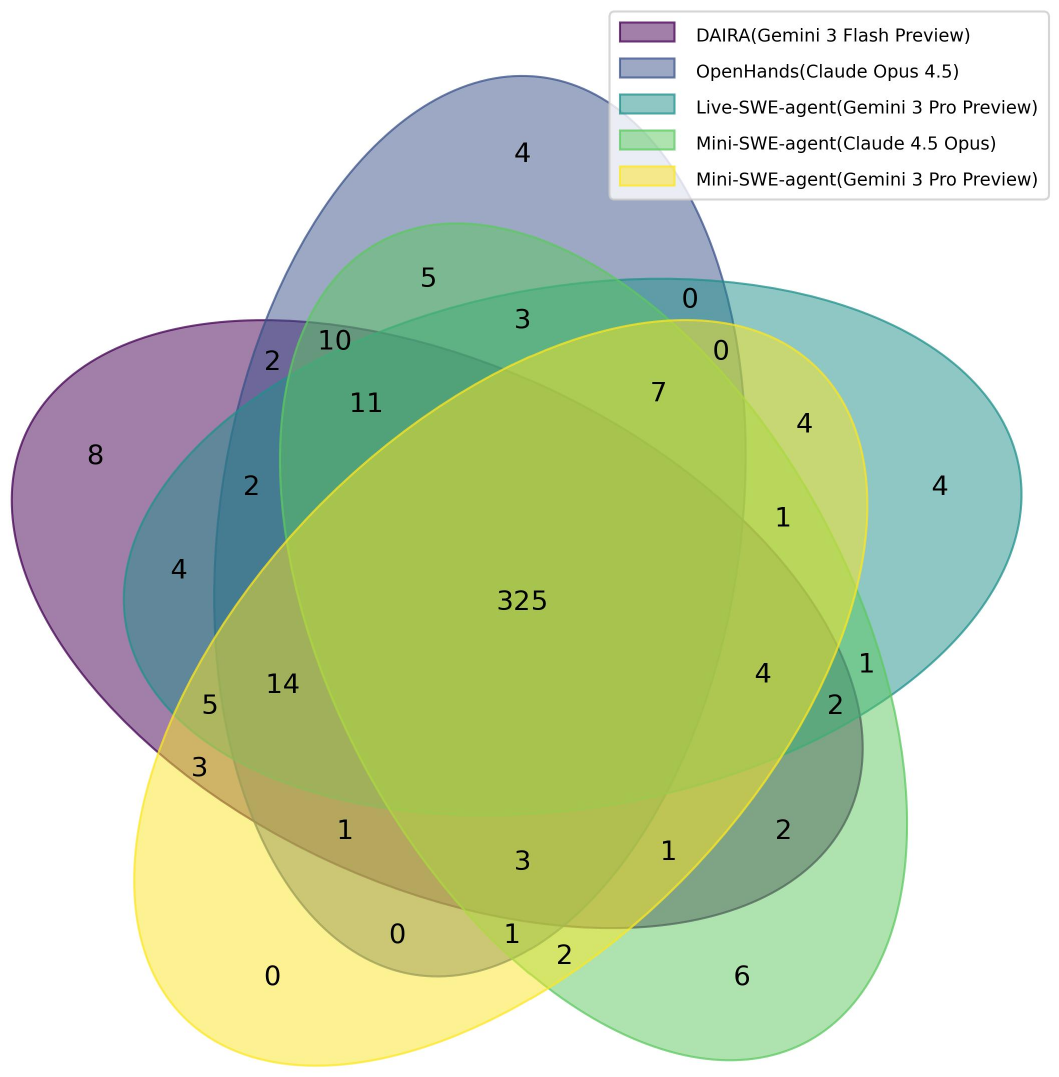}
    \caption{Overlap Analysis of Resolved Instances among the Top-5 Performing Configurations.}
    \label{fig:unique_resolved}
\end{figure}

\textbf{Design.} For RQ1, our objective is to evaluate \ours{}'s effectiveness on real-world issue resolution tasks and compare it with representative repair agents. We consider mainstream SOTA frameworks, including \sweagent, \openhand, \live, and \mini. The resolution rate is used as the primary metric. We report two types of comparisons. First, to contextualize \ours against public SOTA systems, we include the official results reported in their papers, GitHub repositories, or leaderboards, which can be regarded as strong configurations selected by the corresponding systems. These public entries are collected from the SWE-bench Verified leaderboard and official reports of the corresponding systems~\cite{swebench_verified_leaderboard}. Fully re-running all systems under identical backbone models and framework settings is difficult due to regional and temporal restrictions on model access as well as high computational cost. Therefore, to isolate the effect of DAIRA's dynamic-analysis workflow, we conduct controlled reproduction experiments on the \sweagent framework. All controlled runs follow the same common resource configuration used in our evaluation: a maximum of 250 steps per instance, a budget cap of \$1.00, and a sampling temperature of $0$.

\textbf{Result.} As shown in Table~\ref{tab:performance}, \ours performs consistently well across the tested backbones. Notably, with \geminiflash, it achieves a 79.40\% resolution rate, the best among the compared configurations. The framework also improves over \mini (75.8\% with \geminiflash) and the standard \sweagent (73.60\% with \geminiflash). With the \ds backbone, \ours achieves a 74.20\% resolution rate, outperforming vanilla \sweagent's 65.40\%. Figure~\ref{fig:unique_resolved} further illustrates the overlap of resolved instances among the top-performing configurations: 325 instances are resolved by all top configurations, while \ours still resolves 8 instances that the other methods miss. This suggests that structured dynamic evidence helps \ours handle a small set of difficult instances unresolved by the compared methods.

\textbf{Case Study.} To demonstrate the specific advantages that dynamic analysis brings in remediation, we further analyze the resolution trajectories of two issues of different difficulty from our Motivating Example: SymPy-17630 ($>1$ hour) and Matplotlib-22719 ($<15$ min). By examining agent behavioral patterns, we illustrate how dynamic analysis optimizes two critical phases---Fault Localization and Patch Generation---facilitating breakthroughs across different complexity levels.

\textbf{Perspective 1: Behavioral Pattern: From Speculative Attempts to Deterministic Exploration.} In Matplotlib-22719, the baseline agent expended interaction steps (steps 5--49) generating disposable exploratory scripts to probe the \texttt{is\_numlike} and \texttt{convert} functions. Without execution visibility, it had to infer the logic boundary from input-output behavior alone. In contrast, \ours used the execution trace report at step 11 to identify that an empty list incorrectly entered the numerical branch and triggered \texttt{\_api.warn\_deprecated}. This dynamic evidence bridged the cognitive gap, enabling the agent to bypass speculative scripting and pinpoint the root cause directly.

This behavioral shift is also evident in SymPy-17630, where \ours followed an evidence-chain driven exploration that reduced reasoning uncertainty. An initial trace showed that a zero matrix degraded into a scalar 0 during intermediate matrix computation. A second targeted trace then searched further around the degradation point and revealed that the scalar came from the upstream \texttt{MatAdd.doit()} return value, precisely localizing the fault. A third trace then exposed the internal construction logic in \texttt{doit()}, allowing the agent to formulate a verified repair. This progression from data-flow tracing to root-cause discovery turns speculative debugging into a more deterministic process. In contrast, traditional methods remain confined to the reported error location mentioned in the issue and struggle to precisely identify the fault among numerous intermediate execution steps.

Table~\ref{tab:extra_files_explored} further quantifies this localization effect at the trajectory level. Across all tested backbones, \ours explores fewer non-gold files than \sweagent, indicating that dynamic evidence narrows the search space during fault localization.
\begin{table}[htbp]
\centering
\caption{\textbf{Trajectory-level localization efficiency.} Values: avg. non-gold files explored (outside gold patch).}
\label{tab:extra_files_explored}
\renewcommand{\arraystretch}{1.18}
\begin{tabular*}{\columnwidth}{@{\extracolsep{\fill}}lrrl@{}}
\toprule
\textbf{Model} & \textbf{\sweagent} & \textbf{\ours} & \textbf{Change} \\
\midrule
\ds & 53.1 & 33.6 & \scriptsize\textcolor{teal!60!black}{(-36.70\%)} \\
\qwen & 84.6 & 65.1 & \scriptsize\textcolor{teal!60!black}{(-23.00\%)} \\
\geminiflash & 26.5 & 24.6 & \scriptsize\textcolor{teal!60!black}{(-7.20\%)} \\
\bottomrule
\end{tabular*}
\end{table}

\begin{figure}[t] % 1. 鍘绘帀 star (*)锛屼娇鍏朵笉鍐嶉€氭爮鏄剧ず
    \centering
    % 璇风‘淇濆浘鐗囨枃浠跺悕涓庝笂浼犵増鏈竴鑷?    % 2. 灏?\linewidth 鏀逛负 \columnwidth锛岄€傚簲鍗曟爮瀹藉害
    % 濡傛灉鍥剧墖缂╁埌鍗曟爮鍚庝粛鏄惧緱澶ぇ锛屽彲浠ュ皾璇曡涓?0.95\columnwidth
    \includegraphics[width=0.95\columnwidth]{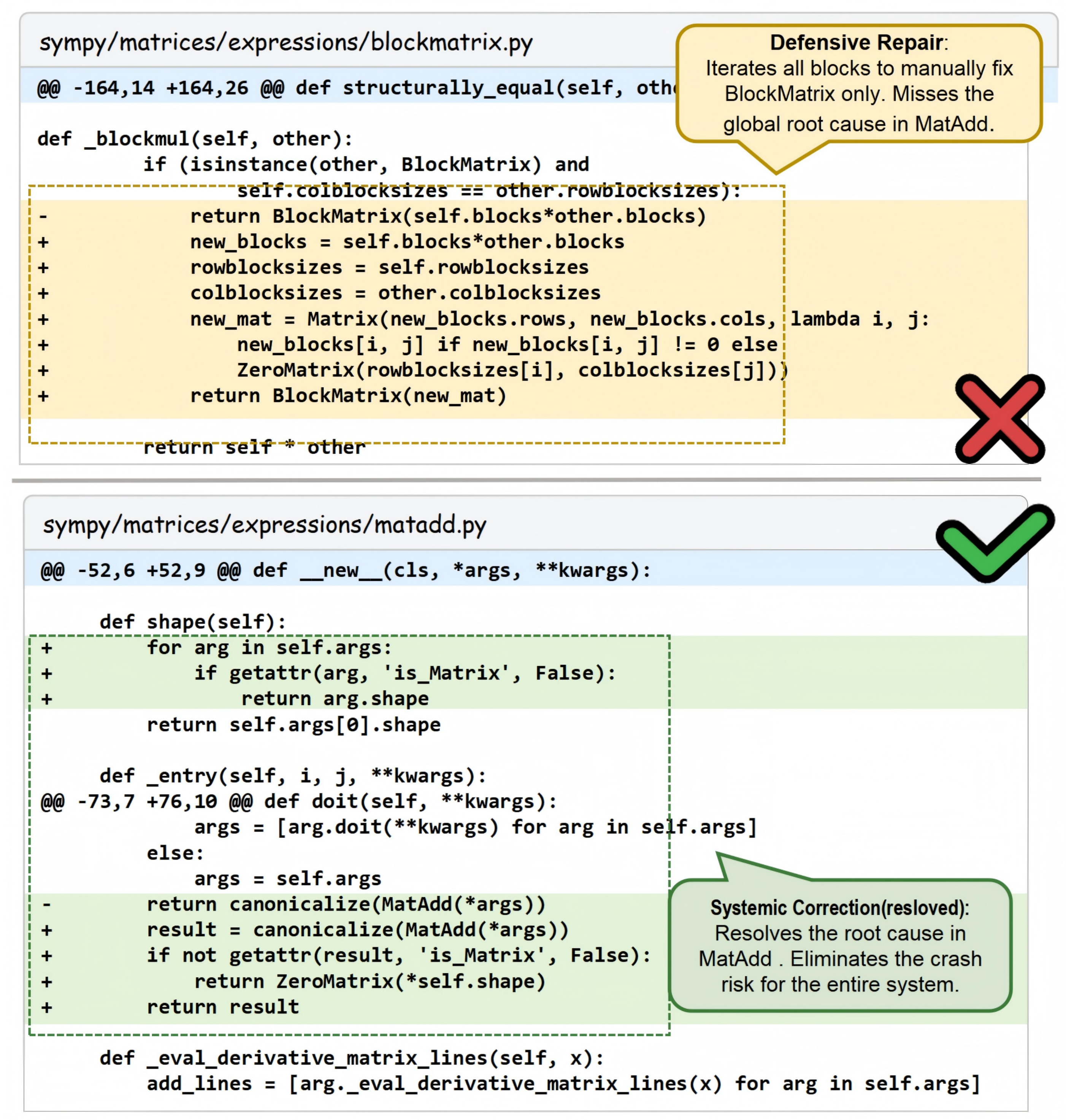}
    \caption{\textbf{Patch Comparison.}}
    \label{fig:patch_comparison}
\end{figure}

\textbf{Perspective 2: Remediation Strategies: From Defensive Repair to Systemic Correction.}
\begin{comment}
    In Matplotlib-22719, where the \sweagent submitted a patch characterized by a strictly defensive strategy. It merely added a size check to suppress the deprecation warning without addressing the completed data conversion process, essentially masking the symptom reported by the user. In contrast, \ours{} demonstrated systemic logical reinforcement. Guided by the full-path execution facts revealed by the execution trace report, the agent refactored the code on three levels: instituting an early exit for empty sequences, adding \texttt{None} filtering within the type determination loop, and rewriting the underlying \texttt{convert\_value} mapping to explicitly handle \texttt{None} as \texttt{np.nan}. This comprehensive repair not only eliminated the warning but also preempted potential system crashes, achieving robust coverage for edge cases.
\end{comment}
In SymPy-17630, the baseline \sweagent applied a defensive repair inside \texttt{blockmatrix.py}, where the final crash was reported. Because it localized the fault to the downstream failure site, its patch retroactively wrapped generated scalar zeros into \texttt{ZeroMatrix} objects, covering the observed crash but leaving the upstream construction logic unchanged. \ours instead followed the trace evidence to the root cause in \texttt{matadd.py}. As shown in Figure~\ref{fig:patch_comparison}, it added matrix-type checking (\texttt{is\_Matrix}) inside \texttt{MatAdd} and ensured that zero-valued matrix expressions return the correct \texttt{ZeroMatrix}. The comparison shows a shift from symptom masking to root-cause repair.

Table~\ref{tab:same_file_patch_quality} further compares patch quality when both agents modify the same files. Under this controlled setting, \ours resolves more instances across all tested backbones, suggesting that dynamic evidence improves not only localization but also the correctness of generated patches.
\begin{table}[htbp]
\centering
\caption{\textbf{Resolution comparison under identical modified files.}}
\small
\label{tab:same_file_patch_quality}
\renewcommand{\arraystretch}{1.18}
\resizebox{\columnwidth}{!}{
\begin{tabular}{lrrc}
\toprule
\textbf{Model} & \textbf{\sweagent} & \textbf{\ours} & \textbf{Relative Improvement} \\
\midrule
DeepSeek V3.2 & 73.17\% & 79.13\% & \scriptsize\textcolor{red}{(+8.15\%)} \\
Gemini 3 Flash Preview & 78.23\% & 80.86\% & \scriptsize\textcolor{red}{(+3.36\%)} \\
Qwen-3-Coder Flash & 53.72\% & 54.44\% & \scriptsize\textcolor{red}{(+1.34\%)} \\
\bottomrule
\end{tabular}
}
\end{table}

\finding{\ours achieves the best overall resolution rate of 79.4\% among the compared configurations and independently resolves 8 additional instances. The case-study tables further show that dynamic evidence reduces non-gold-file exploration and improves patch correctness under identical modified files.}

\subsection{RQ2: Beyond Python}
\textbf{Design.} To examine whether \ours remains effective beyond Python, we evaluate \ours on \swebenchm with \ds. Beyond the Python implementation, we additionally select the C/C++, Java, and Ruby language groups, covering 129 of 300 tasks, and implement language-specific dynamic-analysis backends while preserving the same agent-facing interface and structured trace format. The Python row reports the \ds result on \swebenchv from RQ1.
\begin{table}[htbp]
\centering
\caption{\textbf{Resolution rates across programming languages.}}
\footnotesize
\label{tab:multilingual_generalization}
\renewcommand{\arraystretch}{1.2}
\setlength{\tabcolsep}{4pt}
\resizebox{\columnwidth}{!}{
\begin{tabular}{lrrc}
\toprule
\textbf{Language} & \textbf{\ours} & \textbf{\sweagent} & \textbf{Relative Improvement} \\
\midrule
Python & 74.20\% & 65.40\% & \scriptsize\textcolor{red}{(+13.46\%)} \\
C/C++ & 64.29\% & 57.14\% & \scriptsize\textcolor{red}{(+12.50\%)} \\
Ruby & 63.64\% & 56.82\% & \scriptsize\textcolor{red}{(+12.00\%)} \\
Java & 58.14\% & 51.16\% & \scriptsize\textcolor{red}{(+13.64\%)} \\
\midrule
Average & 71.70\% & 63.28\% & \scriptsize\textcolor{red}{(+13.32\%)} \\
\bottomrule
\end{tabular}
}
\end{table}

\textbf{Result.} As shown in Table~\ref{tab:multilingual_generalization}, \ours improves over \sweagent on Python and all evaluated non-Python language groups, achieving relative improvements of 13.46\%, 12.50\%, 12.00\%, and 13.64\% on Python, C/C++, Ruby, and Java, respectively. These results suggest that structured dynamic execution information is effective beyond the Python ecosystem and can transfer across different language runtimes.

\finding{\ours achieves similar gains on the evaluated non-Python languages, indicating that the proposed workflow transfers beyond Python.}

\subsection{RQ3: Model Impact}
\textbf{Design.} For RQ3, to verify the framework's versatility, we investigate the performance gains and cost implications of \ours across diverse base models. We set up a comparative experiment using three representative models---\ds, \qwen, and \geminiflash---evaluating \ours against the \sweagent baseline for each.
\begin{comment}

\begin{figure*}[t]
  \centering
  % Two-column version kept for reference.
  \subfloat[]{%
    \includegraphics[width=0.5\textwidth]{images_resolution_rate_comparison.pdf}%
    \label{fig:resolution}%
  }%
  \subfloat[]{%
    \includegraphics[width=0.5\textwidth]{images_avg_cost_comparison.pdf}%
    \label{fig:cost}%
  }
  
  \caption{Performance analysis in different model. (a) Resolution Rates. (b) Average Cost.}
  \label{fig:differentmodel}
\end{figure*}

\end{comment}

\begin{figure}[htbp]
  \centering
  
  \subfloat[Resolution Rates.]{%
    \includegraphics[width=0.85\columnwidth]{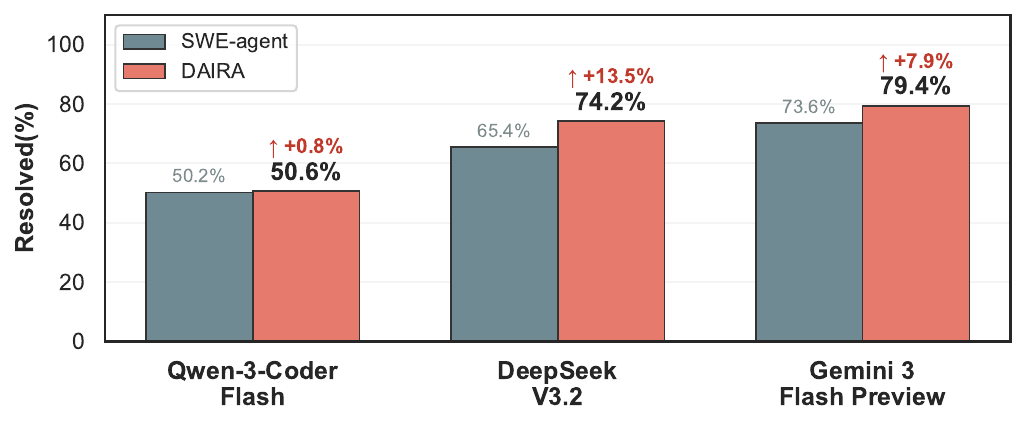}%
    \label{fig:resolution}%
  }
  \vspace{-0.5em}

  \subfloat[Average Cost.]{%
    \includegraphics[width=0.85\columnwidth]{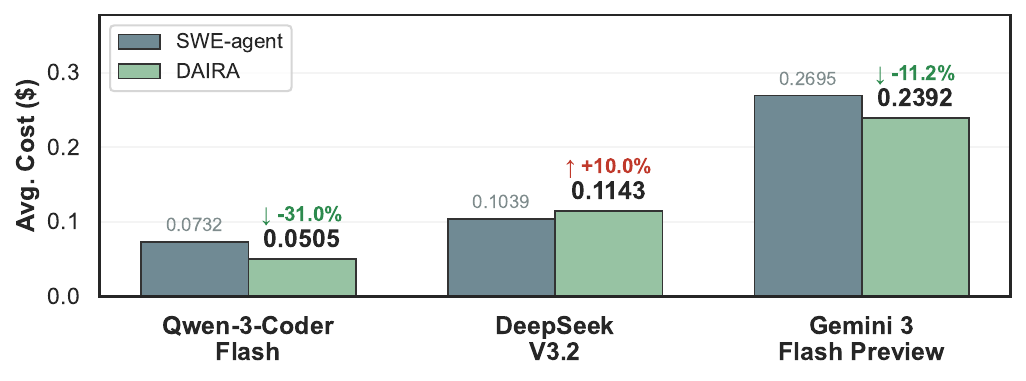}%
    \label{fig:cost}%
  }
  \vspace{-0.4em}
  
  \caption{Performance analysis across different models.}
  \label{fig:differentmodel}
\end{figure}

\textbf{Result.} Figure~\ref{fig:differentmodel} quantifies the effect of \ours on resolution performance and computational cost across different base models. The improvement is most pronounced for stronger models: \ds achieves the largest resolution gain ($+13.5\%$), while \geminiflash improves by $+7.9\%$ and attains the highest absolute resolution rate of 79.4\%. For \qwen, the resolution gain is more limited (+0.8\%), but the overall cost decreases substantially by 31.0\%; \geminiflash also achieves an 11.2\% cost reduction. Together with Table~\ref{tab:extra_files_explored}, these cost reductions indicate that \tracetool can reduce unnecessary retrieval and verification overhead by directing the agent toward more relevant fault locations. In contrast, \ds incurs a 10.0\% cost increase, but this additional expenditure yields the largest resolution gain, indicating a favorable performance-cost trade-off. Overall, \ours remains effective across different backbone capacities, with stronger models converting the added evidence into larger accuracy gains and lighter models gaining more from reduced exploration cost.

\finding{\ours delivers consistent resolution improvements across base models while exhibiting model-dependent cost profiles: it can reduce redundant exploration in cost-sensitive settings and convert additional computation into larger gains on stronger backbones.}

\subsection{RQ4: Robustness}
\textbf{Design.} For RQ4, we evaluate the robustness of \ours across task difficulty levels, with particular attention to high-difficulty issues. We compare its resolution rate with representative baselines within each difficulty level and further analyze the gap against \sweagent across the same strata.

\begin{figure}[htbp]
    \centering
    % 浣跨敤 \linewidth 纭繚鍥剧墖閫傚簲鍗曟爮瀹藉害
    \includegraphics[width=\linewidth]{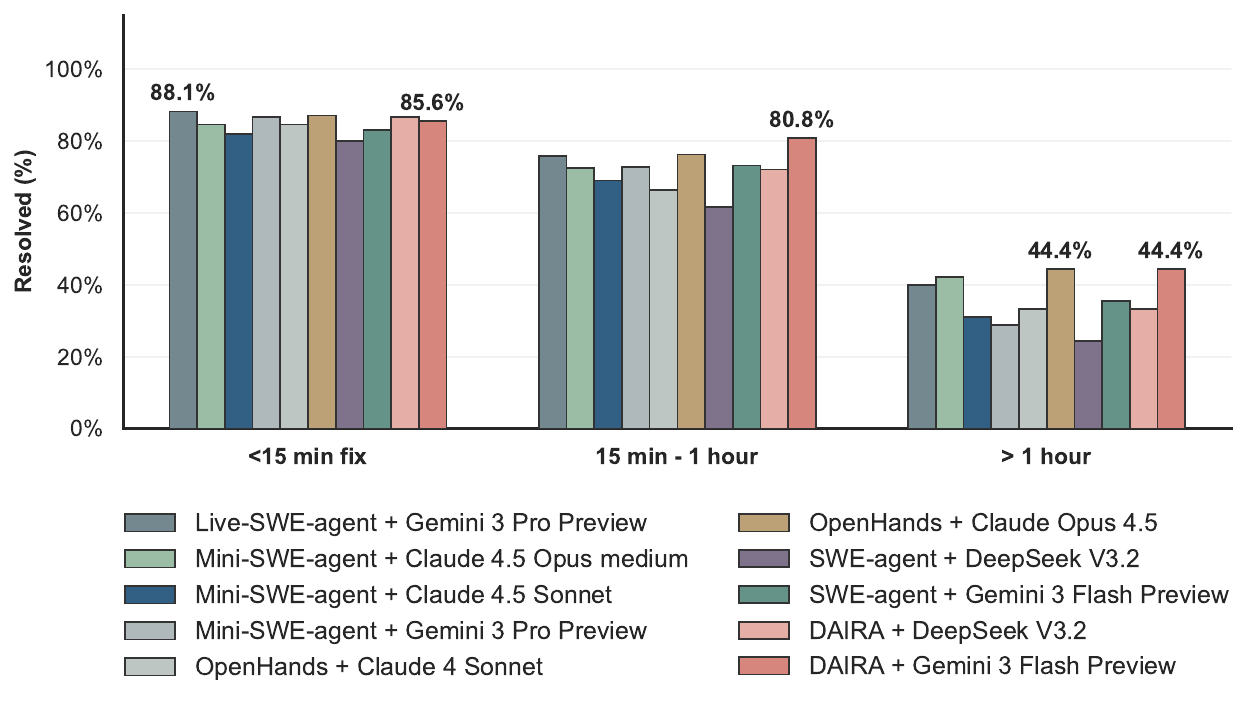}
    \caption{Comparison of Resolution Rates across Different Difficulty Levels.}
    \label{fig:different_method_different_diffculty}
\end{figure}

\textbf{Result.}
%:Scalability to high-difficulty tasks and evolutionary resource allocation.} 
Figure~\ref{fig:different_method_different_diffculty} shows that \ours achieves $80.8\%$ on medium-difficulty tasks and $44.4\%$ on high-difficulty ones. These results indicate that the dynamic information supplied by the dynamic analysis module can help the agent reason about complex runtime-dependent behaviors and improve its ability to resolve more difficult issues.

\textbf{Effectiveness.} As shown in Table~\ref{tab:different_level_resolved_rate}, \ours demonstrates a clear advantage across nearly all difficulty levels. Notably, the performance gain becomes more pronounced as task complexity increases. In the high-difficulty category ($>$1 hour), all models achieved their most substantial relative improvements, with \qwen reaching a remarkable $+133.33\%$ and \ds increasing by $+36.36\%$. This underscores \ours{}'s exceptional capacity for resolving time-consuming, complex tasks. 
The performance drop of \qwen on medium-difficulty tasks primarily reflects instability in its inference when processing dense contexts. Unlike high-performance models, it is more prone to cognitive overload under complex feedback, struggling to consistently extract key information---leading to localized performance fluctuations. Notably, the model's improvement on difficult tasks confirms that \qwen remains effective, with performance constraints stemming from feedback uncertainty rather than a fundamental failure.
\begin{comment}
\begin{table*}[t]
\centering
\caption{\textbf{Performance Comparison by Model and Difficulty.} Note that all reported metrics represent the average value per instance.}
\label{tab:different_level_resolved_rate}
    \begin{tabular}{lllrlr}
    \toprule
    \textbf{Model} & \textbf{Difficulty} & \multicolumn{2}{c}{\textbf{Resolved(\%)}}& \multicolumn{2}{c}{\textbf{Avg Cost (\$)}} \\
    \cmidrule(lr){3-4} \cmidrule(lr){5-6}
    & & \textbf{\ours}& \textbf{\sweagent} & \textbf{\ours} & \textbf{\sweagent}  \\
    \midrule
    \multirow{3}{*}{\textbf{\qwen}} & \textbf{$<$15 min fix} & 71.13\up{(+9.52\%)} & 64.95 & 0.0416\down{(-26.19\%)} & 0.0564 \\
    & \textbf{15 min - 1 hour} & 41.38\down{(-11.48\%)} & 46.74 & 0.0554\down{(-32.32\%)} & 0.0819 \\
    & \textbf{$>$ 1 hour} & 15.56\up{(+133.33\%)} & 6.67 & 0.0604\down{(-36.25\%)} & 0.0947 \\
    \midrule
    \multirow{3}{*}{\textbf{\ds}} & \textbf{$<$15 min fix} & 86.60\up{(+8.39\%)} & 79.90 & 0.0940\up{(+2.81\%)} & 0.0915 \\
    & \textbf{15 min - 1 hour} & 72.03\up{(+16.77\%)} & 61.69 & 0.1209\up{(+10.07\%)} & 0.1098 \\
    & \textbf{$>$ 1 hour} & 33.33\up{(+36.36\%)} & 24.44 & 0.1631\up{(+32.74\%)} & 0.1229 \\
    \midrule
    \multirow{3}{*}{\textbf{\geminiflash}} & \textbf{$<$15 min fix} & 85.57\up{(+3.11\%)} & 82.99 & 0.1929\down{(-13.42\%)} & 0.2228 \\
    & \textbf{15 min - 1 hour} & 80.84\up{(+10.47\%)} & 73.18 & 0.2556\down{(-8.28\%)} & 0.2787 \\
    & \textbf{$>$ 1 hour} & 44.44\up{(+25.00\%)} & 35.56 & 0.3432\down{(-17.78\%)} & 0.4174 \\
    \bottomrule
    \end{tabular}
\end{table*}
\end{comment}
\begin{table}[htbp]
\centering
\caption{\textbf{Performance comparison by model and difficulty.}}
\small
\label{tab:different_level_resolved_rate}
\renewcommand{\arraystretch}{1.28}
\resizebox{\columnwidth}{!}{
    \begin{tabular}{lllr}
    \toprule
    \textbf{Model} & \textbf{Difficulty} & \multicolumn{2}{c}{\textbf{Resolved (\%)}} \\
    \cmidrule(lr){3-4}
    & & \textbf{\ours} & \textbf{\sweagent} \\
    \midrule
    \multirow{3}{*}{\textbf{\qwen}} & \textbf{$<$15 min fix} & 71.13\up{(+9.52\%)} & 64.95 \\
    & \textbf{15 min--1 hour} & 41.38\down{(-11.48\%)} & 46.74 \\
    & \textbf{$>$1 hour} & 15.56\up{(+133.33\%)} & 6.67 \\
    \midrule
    \multirow{3}{*}{\textbf{\ds}} & \textbf{$<$15 min fix} & 86.60\up{(+8.39\%)} & 79.90 \\
    & \textbf{15 min--1 hour} & 72.03\up{(+16.77\%)} & 61.69 \\
    & \textbf{$>$1 hour} & 33.33\up{(+36.36\%)} & 24.44 \\
    \midrule
    \multirow{3}{*}{\textbf{\geminiflash}} & \textbf{$<$15 min fix} & 85.57\up{(+3.11\%)} & 82.99 \\
    & \textbf{15 min--1 hour} & 80.84\up{(+10.47\%)} & 73.18 \\
    & \textbf{$>$1 hour} & 44.44\up{(+25.00\%)} & 35.56 \\
    \bottomrule
    \end{tabular}
}
\end{table}

\begin{table*}[htbp]
\centering
\scriptsize
\caption{\textbf{Comprehensive Performance Comparison.} All metrics are averaged per instance; "Calls" indicates agent iteration steps.}
\label{tab:combine_cost_token_no_cost}
% --- 棰滆壊瀹氫箟 ---
\newcommand{\tabup}[1]{\textcolor{red}{#1}}
\newcommand{\tabdown}[1]{\textcolor{teal!60!black}{#1}}

% --- 鏍煎紡璁剧疆 ---
\renewcommand{\arraystretch}{1.08}
\setlength{\tabcolsep}{2.2pt}

\resizebox{0.90\textwidth}{!}{%
\begin{tabular}{lc|ccc|ccc|ccc}
\toprule
\multirow{2}{*}{\textbf{Model}} & \multirow{2}{*}{\textbf{Method}} & \multicolumn{3}{c|}{\textbf{$<$ 15 min fix}} & \multicolumn{3}{c|}{\textbf{15 min - 1 hour}} & \multicolumn{3}{c}{\textbf{$>$ 1 hour}} \\
\cmidrule(lr){3-5} \cmidrule(lr){6-8} \cmidrule(lr){9-11}
 & & \textbf{Calls} & \textbf{Input} & \textbf{Output} & \textbf{Calls} & \textbf{Input} & \textbf{Output} & \textbf{Calls} & \textbf{Input} & \textbf{Output} \\
\midrule

% --- Qwen Data ---
\multirow{3}{*}{\textbf{\qwen}} & \textbf{\sweagent} & 42.7 & 824k & 9.4k & 51.6 & 1.21M & 12.3k & 56.2 & 1.41M & 13.4k \\
 & \textbf{\ours} & 41.8 & 580k & 10.1k & 49.1 & 784k & 12.2k & 53.6 & 856k & 13.0k \\
 & \textit{Diff} & \tabdown{(-2.1\%)} & \tabdown{(-29.6\%)} & \tabup{(+7.0\%)} & \tabdown{(-4.8\%)} & \tabdown{(-35.2\%)} & \tabdown{(-0.8\%)} & \tabdown{(-4.6\%)} & \tabdown{(-39.1\%)} & \tabdown{(-3.0\%)} \\
\midrule

% --- DeepSeek Data ---
\multirow{3}{*}{\textbf{\ds}} & \textbf{\sweagent} & 89.2 & 2.57M & 21.1k & 97.5 & 3.12M & 24.2k & 103.8 & 3.53M & 27.5k \\
 & \textbf{\ours} & 78.9 & 1.65M & 23.1k & 92.0 & 2.22M & 28.1k & 113.9 & 3.14M & 35.1k \\
 & \textit{Diff} & \tabdown{(-11.5\%)} & \tabdown{(-35.8\%)} & \tabup{(+9.7\%)} & \tabdown{(-5.6\%)} & \tabdown{(-28.9\%)} & \tabup{(+16.5\%)} & \tabup{(+9.8\%)} & \tabdown{(-11.2\%)} & \tabup{(+27.5\%)} \\
\midrule

% --- Gemini Data ---
\multirow{3}{*}{\textbf{\geminiflash}} & \textbf{\sweagent} & 43.0 & 931k & 16.9k & 48.3 & 1.16M & 21.2k & 57.4 & 1.80M & 27.9k \\
 & \textbf{\ours} & 49.6 & 710k & 20.5k & 57.8 & 980k & 24.7k & 71.4 & 1.35M & 31.3k \\
 & \textit{Diff} & \tabup{(+15.3\%)} & \tabdown{(-23.6\%)} & \tabup{(+21.3\%)} & \tabup{(+19.8\%)} & \tabdown{(-15.6\%)} & \tabup{(+16.6\%)} & \tabup{(+24.4\%)} & \tabdown{(-25.3\%)} & \tabup{(+12.4\%)} \\

\bottomrule
\end{tabular}
}
\end{table*}

\textbf{Efficiency.} Table~\ref{tab:combine_cost_token_no_cost} shows an almost consistent reduction in input-token consumption across all models and difficulty levels, even though the dynamic analysis module introduces additional trace context. This indicates that the main efficiency advantage of \ours comes from more precise context retrieval: the agent loads less broad static context while using dynamic execution information to focus subsequent interactions.

Furthermore, distinct resource-allocation patterns emerge across models of varying capabilities. For \qwen, the large input-token reduction ($-29.6\%$ to $-39.1\%$), together with nearly unchanged agent steps and output tokens, shows that \ours curbs lightweight models' reliance on blind, context-heavy trial-and-error without increasing interaction effort. For \geminiflash, agent steps increase ($+15.3\%$ to $+24.4\%$) while input tokens decrease ($-15.6\%$ to $-25.3\%$), indicating a shift from large file loads to more granular queries with smaller per-step contexts.

In contrast, \ds shifts toward deeper exploration as task difficulty increases: its relative increases in agent steps and output tokens grow on harder tasks, with agent steps changing from $-11.5\%$ on easy tasks to $+9.8\%$ on hard tasks, and output tokens rising from $+9.7\%$ to $+27.5\%$. This resource shift corresponds to the largest resolution gain among the tested backbones.

\finding{\ours exhibits strong robustness across varying task difficulties (80.8\% medium, 44.4\% high), yielding peak performance gains in high-difficulty scenarios across all base models. Furthermore, it drives model-specific behavioral shifts to balance performance and computational efficiency.}

\subsection{RQ5: Ablation Study}
\textbf{Design.} To isolate component contributions, we compare the \sweagent baseline, the full \ours framework, and three variants: (1) w/o \tracesummary (uses raw execution traces), (2) w/o \tracetool (removes dynamic analysis module), and (3) w/o \myworkflow (Baseline + Dynamic Analysis module), which integrates the dynamic analysis tool directly into the original \sweagent without the specialized interaction guidance. We evaluate these across resolution rates and costs. We utilize the \ds model across all these evaluations, as it strikes an optimal balance between sufficient capabilities and cost efficiency.

\textbf{Result.} Table~\ref{tab:ablation} details the ablation study of \ours{}. The dynamic analysis module is the primary performance driver. Introducing it yields substantial gains: \ours{} outperforms the variant w/o \tracetool{} ($74.20\%$ vs. $68.20\%$), and adding it to the Baseline improves success from $65.40\%$ to $70.20\%$. This highlights that runtime feedback crucially bridges the information gap inherent in relying solely on static analysis.
\begin{comment}
\begin{figure*}[t]
  \centering
  \normalfont %
  \caption{Ablation Study of \ours. Note that ``$w/o$ \myworkflow'' represents a variant adding the complete dynamic analysis module directly to the \sweagent.}
  \label{tab:ablation}
  \renewcommand{\arraystretch}{1.1} 
  \begin{tabular}{llll}
    \toprule
    \textbf{Experiment} & \textbf{Success Rate} & \textbf{Avg. Cost(\$)} & \textbf{Avg. Calls} \\
    \midrule
    \textbf{Baseline(\sweagent)}  & 65.40\% & 0.7336 & 94.82 \\
    \midrule
    \textbf{\ours(ours) } & \textbf{74.20\%} & 0.8114 & 88.89 \\
    \textbf{w/o \tracesummary}  & 65.80\%\down{(-11.32\%)} & 0.8727\up{(+7.56\%)} & \textbf{82.30}\down{(-7.41\%)} \\
    \textbf{w/o \tracetool}  & 68.20\%\down{(-8.09\%)} & \textbf{0.7057}\down{(-13.03\%)} & 83.92\down{(-5.59\%)} \\
    \textbf{w/o \myworkflow}  & 70.20\%\down{(-5.39\%)} & 0.9776\up{(+20.48\%)} & 109.64\up{(+23.34\%)} \\
    \bottomrule
  \end{tabular}
\end{figure*}
\end{comment}
\begin{table}[htbp]
  \centering
  \caption{Ablation study of \ours.}  
  \label{tab:ablation}
  \renewcommand{\arraystretch}{1.08}
  \scriptsize
  \setlength{\tabcolsep}{2pt}
  \newcommand{\diffup}[1]{\textcolor{red}{#1}}
  \newcommand{\diffdown}[1]{\textcolor{teal!60!black}{#1}}
  \begin{tabular}{@{}p{0.60\columnwidth}>{\raggedleft\arraybackslash}p{0.18\columnwidth}>{\raggedleft\arraybackslash}p{0.18\columnwidth}@{}}
    \toprule
    \textbf{Experiment} & \multicolumn{1}{c}{\textbf{Success Rate}} & \multicolumn{1}{c@{}}{\textbf{Avg. Cost (\$)}} \\
    \midrule
    \textbf{Baseline (\sweagent)} & 65.40\% & 0.7336 \\
    \midrule
    \textbf{\ours} & \textbf{74.20\%} & 0.8114 \\
    \midrule
    \textbf{w/o \tracesummary} & 65.80\% & 0.8727 \\
    \textit{Diff} & \diffdown{(-11.32\%)} & \diffup{(+7.56\%)} \\
    \textbf{w/o \tracetool} & 68.20\% & \textbf{0.7057} \\
    \textit{Diff} & \diffdown{(-8.09\%)} & \diffdown{(-13.03\%)} \\
    \textbf{w/o \myworkflow} & 70.20\% & 0.9776 \\
    \textit{Diff} & \diffdown{(-5.39\%)} & \diffup{(+20.48\%)} \\
    \bottomrule
  \end{tabular}
\end{table}

\noindent The \tracesummary{} module is equally vital. Removing it plummets the success rate to $65.80\%$, even with raw execution traces retained. This exposes the volatility of raw execution traces: directly injecting them into the context window introduces severe noise that degrades reasoning. Standardizing these traces into structured summaries is imperative to unlock the true potential of dynamic analysis.
Finally, directly integrating dynamic analysis (i.e., Baseline + Dynamic Analysis) improves performance but incurs the highest cost ($\$0.9776$) and API calls ($109.64$) due to unguided exploration. Conversely, \myworkflow{} optimizes the interaction logic for dynamic feedback. It achieves the highest overall success rate while reducing costs by $17\%$ ($\$0.8114$) compared to direct integration, striking an optimal balance between performance and efficiency.

\finding{\ours achieves an optimal balance between performance and efficiency through the synergistic integration of the \tracetool, \tracesummary, and \myworkflow.}

\section{Related work}\label{sec:Relatework}
\textbf{Static-analysis-based Approaches for Issue Resolution.}
Static analysis approaches~\cite{xia2024agentlessdemystifyingllmbasedsoftware,zhang2024autocoderoverautonomousprogramimprovement,ruan2024specrovercodeintentextraction,yang2025enhancingrepositorylevelsoftwarerepair,liu2024codexgraphbridginglargelanguage} mitigate LLM context constraints in repository-level repair by deterministically extracting critical code structures to retrieve necessary context. For instance, \textbf{Agentless}~\cite{xia2024agentlessdemystifyingllmbasedsoftware} demonstrates that static retrieval enables high-performance patch generation by decoupling fault localization from repair. To refine retrieval precision, \textbf{AutoCodeRover}~\cite{zhang2024autocoderoverautonomousprogramimprovement} and \textbf{SpecRover}~\cite{ruan2024specrovercodeintentextraction} leverage ASTs and intent analysis, respectively, while \textbf{KGCompass}~\cite{yang2025enhancingrepositorylevelsoftwarerepair} applies knowledge graphs to enforce global dependency constraints. Nevertheless, relying exclusively on static snapshots obscures intermediate runtime states. Our work addresses this critical limitation by integrating dynamic analysis to supply concrete execution evidence.

\textbf{Agentic Approaches for Issue Resolution.} Agent-based approaches~\cite{NEURIPS2024_5a7c9475,antoniades2025swesearchenhancingsoftwareagents,mu2025experepairdualmemoryenhancedllmbased,wang2025openhandsopenplatformai,li-etal-2025-codetree,li2025swedebatecompetitivemultiagentdebate,chen2025sweexpexperiencedrivensoftwareissue,xia2025livesweagentsoftwareengineeringagents} introduce loop architectures that dynamically perceive feedback and adjust actions. Pioneered by \textbf{\sweagent}~\cite{NEURIPS2024_5a7c9475} via a specialized ACI, recent methods have rapidly evolved to address specific reasoning and execution limitations. Specifically, to mitigate greedy reasoning, \textbf{SWE-search}~\cite{antoniades2025swesearchenhancingsoftwareagents} and \textbf{CodeTree}~\cite{li-etal-2025-codetree} integrate tree-search algorithms; to overcome memory constraints, \textbf{SWE-Exp}~\cite{chen2025sweexpexperiencedrivensoftwareissue} and \textbf{EXPEREPAIR}~\cite{mu2025experepairdualmemoryenhancedllmbased} leverage historical repair data; while \textbf{\live}~\cite{xia2025livesweagentsoftwareengineeringagents} enables real-time self-evolution. However, these agents mainly rely on superficial execution input-output feedback. They fundamentally lack the intermediate execution-level observability crucial for diagnosing the internal mechanisms of complex defects.

\section{Threats to validity}\label{sec:Threat2validity}
\textbf{Internal Validity.} We mitigate data leakage by evaluating on the curated \swebenchv and \swebenchm. We compare \ours with \sweagent using the same backbones and configuration, reducing the risk that gains stem from model memorization.

\textbf{External Validity.} We implement and evaluate \ours across Python, C/C++, Ruby, and Java, suggesting that its benefits are not tied to a single language ecosystem. Broader evaluations across more languages, runtimes, and bug types remain necessary; however, mature debugging infrastructure makes extension mainly a matter of replacing the underlying instrumentation backend.

\section{Conclusion}\label{sec:conclusion}
We present \ours, an agentic framework that integrates dynamic analysis into repository-level issue resolution. By converting execution traces into structured reports, \ours provides agents with dynamic execution information that complements static code exploration and helps reduce speculative search. Evaluations on \swebenchv show that \ours achieves the best overall resolution rate among the compared configurations, improves trajectory-level localization efficiency, and remains effective on complex, high-difficulty tasks. Similar gains are also observed on other languages in \swebenchm. Overall, our findings show that dynamic analysis is a practical way to make autonomous repair agents evidence-guided and efficient.

\section{DATA AVAILABILITY}
Our data and code are available at: \url{https://anonymous.4open.science/r/DAIRA-80AC}.

\bibliographystyle{IEEEtran}
\bibliography{bib}

\end{document}